\documentclass[12pt]{article} 
\usepackage{epsf}

\begin{document}

\renewcommand{\theequation}{\arabic{section}.\arabic{equation}} 
\renewcommand{\thefootnote}{\fnsymbol{footnote}}                
\newfont{\elevenmib}{cmmib10 scaled\magstep1}
\newfont{\cmssbx}{cmssbx10 scaled\magstep3}                     
\newcommand{\polya}{P\'{o}lya }                                 
\newcommand{\preprint}{                                         
            \begin{flushleft}                                   
            \elevenmib Yukawa\, Institute\, Kyoto               
            \end{flushleft}\vspace{-1.3cm}                      
            \begin{flushright}\normalsize  \sf                  
            quant-ph/9611047              
            \\November 1996                                     
            \end{flushright}}                                   
\newcommand{\Title}[1]{{\baselineskip=26pt \begin{center}       
            \cmssbx                               
             #1 \\ \ \\ \end{center}}}                
\newcommand{\FU}{\begin{center}\large \bf                       
            Hong-Chen Fu\footnote[1]{On leave of absence from   
            Institute of Theoretical Physics, Northeast         
            Normal University, Changchun 130024, P.R.China.     
            E-mail: hcfu@yukawa.kyoto-u.ac.jp }\end{center}}    
\newcommand{\YITP}{\begin{center} \it                           
            Yukawa Institute for Theoretical Physics, Kyoto     
            University,\\ Kyoto 606-01, Japan \end{center}}     
\newcommand{\Accepted}[1]{\begin{center}{\large \sf #1}\\       
            \vspace{1mm}{\small \sf Accepted for Publication}   
            \end{center}}                                       
\baselineskip=20pt
\preprint\bigskip
\Title{  \polya States of Quantized Radiation Fields, their Algebraic
         Characterization and Nonclassical Properties}

\FU
\YITP

\vspace{3.3cm}

\begin{abstract}
\polya states of single mode radiation field are proposed and
their algebraic characterization and nonclassical properties
are investigated. They degenerate to the binomial (atomic
coherent) and negative binomial (Perelomov's su(1,1) coherent) 
states in two different limits and further to the number, the 
ordinary coherent and Susskind-Glogower phase states. The
algebra involved turn out to be a two-parameter deformation
of both su(2) and su(1,1). Nonclassical properties are 
investigated in detail.
\end{abstract}

\newpage

\section{Introduction}
\setcounter{equation}{0}

Since Stoler {\it et al.} introduced the binomial states in
1985 \cite{stol}, the so-called {\it intermediate states} 
have attracted attentions. An important feature of these 
states is that they interpolate between two fundamental
states, such as the number, the coherent and squeezed and
the phase states, and reduce to them in two different 
limits. For example, the binomial states (BS) \cite{stol,barr}
between the number and the coherent states; the negative 
binomial states (NBS) \cite{nbs} between the coherent and 
the Susskind-Glogower (SG) phase states \cite{sg}; the 
hypergeometric states (HGS) between the number and the coherent 
states \cite{hgs}; the intermediate number-squeezed states 
\cite{ins1,ins2} and the intermediate number-phase states 
\cite{inp}. Another feature of some intermediate states is 
that their photon distributions are some famous probability 
distributions in probability theory. BS corresponds to the 
binomial distribution \cite{stol}, NBS to the negative 
binomial distribution \cite{nbs} and HGS to the hypergeometric 
distribution \cite{hgs}.

In this letter we shall introduce the {\it P\'{o}lya states} 
(PS) in the same way as the BS from the binomial distribution
\cite{pro}, namely, we define the {\it P\'{o}lya states} as
{\em probability amplitudes} of the {\it P\'{o}lya} distribution.
We find that, as intermediate states, 
PS interpolate between the BS and NBS, or in other words, the 
atomic coherent states and the Perelomov's su(1,1) coherent 
states. Furthermore, the PS tend to the number and the coherent 
states (from BS) and the coherent and the SG phase states (from
NBS). So the present letter supplies a unified approach to these
important quantum states in quantum optics. As in the cases of 
BS and NBS, the PS also admit the ladder-operator formalism and 
the algebra involved is a two-parameter deformation of 
Holstein-Primakoff (HP) realization of both su(2) and su(1,1) 
in the sense that it contracts to their universal enveloping 
algebras in two different limits. As far as I know, this kind 
of deformed algebras has not appeared in the literature. The 
nonclassical properties of PS are also investigated. The field 
in PS is of sub-Poissonian and squeezed in some ranges of 
parameters involved.   

\section{P\'{o}lya states and their limiting states}
\setcounter{equation}{0}
\newcommand{\be}{\begin{equation}}
\newcommand{\ee}{\end{equation}}
\newcommand{\ga}{\gamma}
\newcommand{\rr}{\rangle}

We define the P\'{o}lya states as
\be
   |M,\ga,\eta\rr=\sum_{n=0}^{M} [P_n^M(\ga,\eta)]^{1\over 2}
   |n\rr,
\ee
where $|n\rr$ is the number state of a single mode radiation field
\begin{equation}
   [a,\,a^{\dagger}]=1,\ \ \ N\equiv a^{\dagger}a,\ \ \
   a|0\rr=N|0\rr=0,\ \ \            
   |n\rr=\frac{a^{\dagger n}}{\sqrt{n!}}|0\rr.
\end{equation}
$M$ is a positive integer, $\ga>0$ is a real constant, $\eta$
is the probability satisfying $0<\eta<1$ and the {\it photon
distribution} $|\langle n|M,\ga,\eta\rr|^2\equiv P_n^M(\ga,\eta)
$ is the P\'{o}lya distribution in probability
theory ($\bar{\eta}=1-\eta$) \cite{pro}
\be
   P_n^M(\ga,\eta)=\left(\begin{array}{c}M\\n\end{array}\right)
   \frac{\eta(\eta+\ga)\cdots (\eta+(n-1)\ga)\,
         \bar{\eta}(\bar{\eta}+\ga)\cdots (\bar{\eta}+(M-n-1)\ga)}{
         (1+\ga)(1+2\ga)\cdots (1+(M-1)\ga)}. \label{polya} 
\ee
The \polya states defined above is obviously normalized since as a
probability distribution $P_n^M(\ga,\eta)$ satisfies
$\sum_{n=0}^{M}P_n^M(\ga,\eta)=1$.

It is well known that the \polya distribution tends to the binomial
and negative binomial distributions in the limit $\gamma\to 0$ (
called the {\em BS limit}, for convenience) and
$M\to\infty, \ga\to 0, \eta\to 0$ with $M\eta=\lambda$ and 
$M\ga=\rho^{-1}$ (called the {\em NBS limit}), respectively
\cite{pro}
\[
P_n^M\to \left\{ \begin{array}{ll}
        \left(\begin{array}{c}M\\ n\end{array}\right)\eta^n
        (1-\eta)^{M-n},
     &  \mbox{in the BS limit,}\\ 
        \left(\begin{array}{c}\lambda\rho+n-1\\ n\end{array}
        \right) \left(\displaystyle 1-\frac{1}{1+\rho}
        \right)^{\lambda\rho}\left(\displaystyle \frac{1}{1+\rho}
        \right)^n,
     & \mbox{in the NBS limit.}
\end{array} \right.
\]
Accordingly the PS go to the BS and NBS in the BS and NBS limits, 
respectively. Furthermore, the BS degenerate to the number and 
coherent states in two different limits \cite{stol} and the NBS 
to the coherent and SG phase states in two different limits
\cite{nbs}. So the PS include the number, the coherent states 
and SG phase states as their limiting states. Therefore, the 
PS interpolate between the BS and NBS, or in other words, 
between the atomic coherent states and Perelomov's su(1,1) 
coherent states. 

\section{Algebraic Characterization}
\setcounter{equation}{0}

Both BS and NBS admit the ladder-operator description, namely, they
satisfy the eigenvalue equations of generators of su(2) or
su(1,1), respectively. In fact, the PS admit the ladder-operator
description, too. It is easy to verify that PS satisfy the following
eigenvalue equation
\begin{equation}
   \gamma\left[(M-N)\left(\frac{\bar{\eta}}{\gamma}+M-N-1\right)
   \left(\frac{\eta}{\gamma}+N\right)\right]^{1\over 2}a|M,\gamma,
   \eta\rr=\gamma(M-N)\left(\frac{\eta}{\gamma}+N\right)|M,\gamma,
   \eta\rr. \label{eigenps}
\end{equation}
Then in the BS or NBS limits, (\ref{eigenps}) tend to
the ladder-operator forms of BS and NBS
\begin{eqnarray}
   &&\sqrt{1-\eta}J^-_M|M,0,\eta\rr=\sqrt{\eta}(M-N)|M,0,\eta\rr,\ \ \ 
     J^-_M\equiv \sqrt{M-N}\,a,   \label{lims1}\\
   &&\sqrt{\rho+1} K^-_{\lambda\rho}|\infty,0,0\rr=(\lambda\rho+N)
     |\infty,0,0\rr,     \ \ \
     K^-_{\lambda\rho}\equiv \sqrt{\lambda\rho+N}\,a,
     \label{lims2}
\end{eqnarray}
where $J^-_M$ and $K^-_{\lambda\rho}$ are the lowering operators
of su(2) and su(1,1) algebras via their HP realizations.
Both limiting results (\ref{lims1}, \ref{lims2}) suggest us defining
the operator on the left side of (\ref{eigenps}) as the {\it lowering 
operator} (up to a constant) of the algebra related to PS
\begin{equation}
   A^-=\frac{\gamma}{\sqrt{(1-\eta)(1+M\gamma)(M\gamma+\eta)}}
   \left[(M-N)\left(\frac{\bar{\eta}}{\gamma}+M-N-1\right)
   \left(\frac{\eta}{\gamma}+N\right)\right]^{1\over 2}a.
\end{equation}
Then the algebraic relations among $A^-$, the {\it raising operator}
$A^+\equiv (A^-)^\dagger$ and $N$ are obtained as
\begin{equation}
   [N,\,A^{\pm}]=\pm A^{\pm}, \ \ \ \
   A^+A^-=F(N), \ \ \ \
   A^-A^+=F(N+1),
\end{equation}
where $F(N)$ is a non-negative hermitian function
\begin{equation}
   F(N)=\frac{N(M-N+1)(\bar{\eta}+\gamma M-\gamma N)(\eta+\gamma N
        -\gamma)}{(1-\eta)(\gamma M+1)(\gamma M+\eta)}.
\end{equation}
This means that the related algebra, which is an associative algebra
generated by $A^-$, $A^+$, $N$ and the unit 1, is a {\it generally deformed
oscillator} with the {\it structure function} $F(N)$. This algebra
has an $(M+1)$-dimensional representation on the Fock space because
of the condition $A^+|M\rr=F(M+1)|M+1\rr=0$.

A remarkable feature of this algebra is that in the BS and NBS limits
it contracts to the universal enveloping algebras of compact su(2) 
and noncompact su(1,1) Lie algebra:
\begin{equation}
   A^- \longrightarrow\left\{\begin{array}{ll}
   \sqrt{M-N}a\equiv J^-_M  & \mbox{in the BS limit,}\\
   \sqrt{\lambda\rho+N}a\equiv K^-_{\lambda\rho} &
   \mbox{in the NBS limit.}
   \end{array}\right.
\end{equation}
Accordingly, its finite dimensional representation degenerates to
a finite dimensional irreducible representation of su(2) with the
highest weight $M/2$ and the infinite dimensional irreducible 
positive discrete representation of su(1,1) with the Bargmann index
$\lambda\rho/2$.

\section{Nonclassical Properties}
\setcounter{equation}{0}

\subsection{Photon Statistics}

The averages $\langle N\rr$, $\langle N^2\rr$ and fluctuation
$\langle \Delta N^2\rr$ are obtained as
\be
       \langle N\rr=M\eta,   \ \ \
       \langle N^2\rr=M\eta+\frac{M\eta(M-1)(\eta+\ga)}{1+\ga},
               \ \ \
       \langle \Delta N^2\rr=\frac{M\eta(M\eta+1)(1-\eta)}
               {1+\ga},
\ee
Then we can easily derive Mandel's $Q$-factor
\be
       Q^M_\gamma(\eta)=\frac{\langle \Delta N^2\rr-
         \langle N\rr}{\langle N\rr}
        =\frac{(M-1)\gamma}{1+\gamma}-\eta\frac{M\ga+1}{1+\gamma},
\ee
which is obviously a linear function of $\eta$ and is a
straight line (we call it the $Q$-line for convenience)
connecting the point $\left(0, Q_\ga^M(0)\right)$ and
$\left(1, Q_\ga^M(1)\right)$, where
\be
       Q_\ga^M(0)=\frac{(M-1)\gamma}{1+\gamma}\equiv
                  (M-1)\left(1-\frac{1}{1+\gamma}\right),\ \ \ \
       Q_\ga^M(1)=-1,
\ee
as illustrated in the Figure 1. We find that

1. In the case $M=1$ or $\gamma=0$ (BS limit),
we have $Q_\ga^M(0)=0$ and the $Q$-line connects (0,0) and
$(1,-1)$. So in this case $Q_\gamma^M(\eta)=-\eta<0$ and the
field is of sub-Poissonian character except for $\eta=0$, which 
corresponds to the Poissonian statistics.

2. If $\gamma>0$ and $M\neq 1$, then $Q_\ga^M(0)>0$ and the $Q$-line must
intersect with the line $Q_\ga^M(\eta)=0$ at the point
\be
     \left(\frac{(M-1)\gamma}{M\ga+1},\, 0\right)
\ee
(see Fig.1). This means that, when
$\eta>\frac{(M-1)\gamma}{M\gamma+1}$ (or
$\eta<\frac{(M-1)\gamma}{M\gamma+1}$ ), $Q_\gamma^M(\eta)<0$
(or $Q_\ga^M(\eta)>0$) and the field in PS is of sub-Poissonian
(super-Poissonian). The point $\eta=\frac{(M-1)\gamma}{M\gamma+1}$
corresponds to the Poissonian statistics. In this case,
the value of $M$ and $\eta$
will affect the ranges of sub-Poissonian (or super-Poissonian)
statistics. The larger $M$ or/and $\gamma$, the larger
$Q_\gamma^M(0$) and therefore $\frac{(M-1)\gamma}{M\ga+1}$.
So the sub-Poissonian range $\frac{(M-1)\gamma}
{M\ga+1}<\eta<1$ becomes smaller.

\subsection{Squeezing Effect}

It is easy to evaluate that
\be
       a^k|M,\ga,\eta\rr=\left[\prod_{i=0}^{k-1}
       (M-i)\frac{k\gamma+\eta}{k\gamma+1}\right]^{1\over 2}
       \left| M-k,\frac{\gamma}{k\gamma+1},
       \frac{k\gamma+\eta}{k\gamma+1}\right\rr,
\ee
for $k\leq M$ and $a^k|M,\gamma,\eta\rangle=0$ for $k>M$.
Define the coordinate $x$ and the momentum $p$ as
\be
       x=\frac{1}{\sqrt{2}}(a^\dagger+a),\ \ \ \
       p=\frac{i}{\sqrt{2}}(a^\dagger-a).
\ee
Then their variances are obtained as
\begin{eqnarray}
  \langle\Delta x^2\rr&=&\frac{1}{2}+M\eta+
       \left[{M\eta(M-1)\frac{\eta+\gamma}{\gamma+1}}\right]^
       {1\over 2}\sum_{n=0}^{M-2}
       \sqrt{P_n^M(\ga,\eta) P_n^{M-2}\left(\frac{\gamma}
       {2\gamma+1},\frac{2\gamma+\eta}{2\gamma+1}\right)}\nonumber \\
       &&-2M\eta\left[\sum_{n=0}^{M-1}\sqrt{P_n^M(\ga,\eta)
       P_n^{M-1}\left(\frac{\gamma}{\gamma+1},\frac{\gamma+\eta}
       {\gamma+1}\right)}\right]^2, \\
  \langle\Delta p^2\rr&=&\frac{1}{2}+M\eta-
       \left[{M\eta(M-1)\frac{\eta+\gamma}{\gamma+1}}\right]^
       {1\over 2}\sum_{n=0}^{M-2}
       \sqrt{P_n^M(\ga,\eta) P_n^{M-2}\left(\frac{\gamma}
       {2\gamma+1},\frac{2\gamma+\eta}{2\gamma+1}\right)}.
\end{eqnarray}

Figures 2 and 3 are plots showing how $\langle \Delta x^2\rangle$
and $\langle \Delta p^2\rangle$ depend on the parameter $\gamma$ 
and $\eta$, respectively. In each case,  different values of
$M$ (5 and 20) are chosen. From these plots we find that:

{\it Quadrature $x$} (see Fig.2). When $\gamma=0$ (BS case), 
the quadrature $x$ is squeezed
in a considerable range $0<\eta \leq \eta_0<1$ of values of $\eta$,
with a maximum of squeezing (minimum of $\langle \Delta x^2\rangle$
that depends on $M$ (the larger $M$, the wider the range and the
smaller $\langle \Delta x^2\rangle$), as indicated in \cite{stol}
and Fig.2. With the increase of $\gamma$, the squeezing range
becomes smaller and smaller and $\langle \Delta x^2\rangle$ 
becomes larger and larger until the squeezing disappears for 
large enough $\gamma$. For large $M$, the 
squeezing disappears faster than that for small $M$.

{\it Quadrature $p$} (see Fig.3). It is well known
that there is no squeezing
for $\gamma=0$ (BS). However, with the increase of $\gamma$,
the quadrature $p$ becomes squeezed drastically in the range of 
$0<\eta\leq\eta_0<1$ and $|\langle \Delta p^2\rangle|$ decreases
drastically until the maximum of squeezing is reached. Then, by
further increasing $\gamma$, the squeezing range becomes smaller 
and smaller and squeezing becomes weaker and weaker. However, the
quadrature $p$ is still squeezed for a very large value of 
$\gamma$. In fact, we can check that only when $\gamma\to\infty$,
$\langle \Delta p^2\rangle$ goes to $1/2$. We also see that
$\langle \Delta p^2\rangle$ for large $M$ is more sensitive
to the parameter $\gamma$ than that for small $M$.

\section{Conclusion}

In this letter we have introduced and investigated the
\polya states and found that:

1. As intermediate states, the \polya states interpolate
the binomial states (or the atomic states) and the
negative binomial states (or the Perelomov's coherent states).

2. Ladder-operator forms of BS and NBS, which are related to 
su(2) and su(1,1) algebras respectively, are generalized to
the PS case. This algebraic characterization leads to an
algebra which is a two-parameter ($\eta$ and $\gamma$) 
deformation of universal enveloping algebra of Lie algebras
su(2) and su(1,1) and contracts to them in two different
limits. This is natural since the PS is an intermediate
state between su(2) (BS) and su(1,1) (NBS) coherent states.
To our knowledge this kind of algebras which mixes su(2) and its
noncompact counterpart su(1,1) has not appeared before in the
literature.

3. We have indicated in \cite{nbs} that the nonclassical
properties of BS and NBS are complementary. As states
interpolating the BS and NBS the PS clearly share the
characters of both BS and NBS: the field in PS is of 
sub-Poissonian character in some range of parameters involved 
and of super-Poissonian character in a different region of 
parameters, and both quadratures $x$ and $p$ are
squeezed in considerable ranges of parameters.

\section*{Acknowledgments}

The author thanks Prof.\,Ryu Sasaki for valuable discussions 
and comments. He is grateful to Japan Society for the Promotion 
of Science (JSPS) for the fellowship. This work is also supported 
in part by the National Science Foundation of China.

          
\section*{Appendix: The P\'{o}lya distribution}

P\'{o}lya originally introduced the  P\'{o}lya distribution 
in 1930 \cite{pol} when considering the sampling from a finite population 
of objects, the numbers of which change with the removal of 
each individual unit. Suppose an urn contains $a$ white balls 
and $b$ black balls. A ball is chosen at random and replaced, 
together with $c$ balls of the same kind. If $M$ successive 
drawing have already been made, of which $n$ are white and 
$M-n$ black, the probability $P_n$ of obtaining $n$ white
balls in a sequence of $M$ is
\[
   P_n=\left(\!\!\!\begin{array}{c}M\\n\end{array}\!\!\!\right)
       \frac{a(a+c)\cdots [a+c(n-1)]b
       (b+c)\cdots [b+c(M-n-1)]}{(a+b)(a+b+c)\cdots
       [a+b+c(M-1)]}.
\]
This is just the  P\'{o}lya distribution (\ref{polya}) if we put
\[
  \eta=\frac{a}{a+b},\ \ \ \ \   
  \bar{\eta}=\frac{b}{a+b}, \ \ \ \ \
  \gamma=\frac{c}{a+b}.
\]
For more details please see \cite{pro}.


\newpage

\begin{figure}
\centerline{\epsfxsize=8cm 
\epsfbox{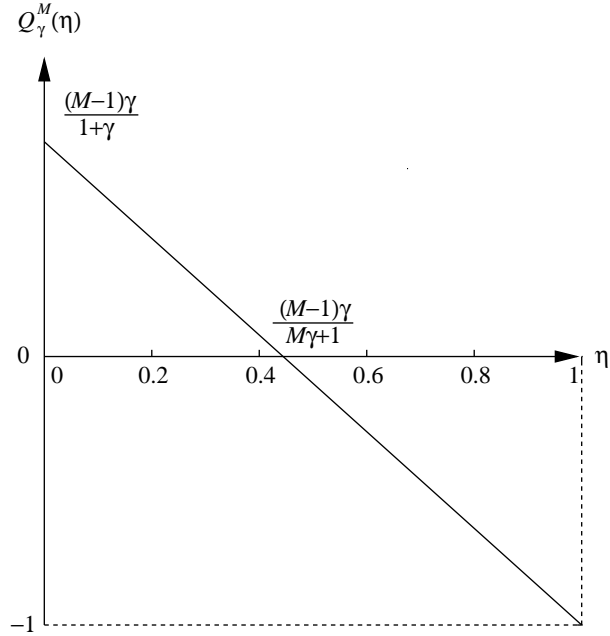}}
\caption{The Mandel's $Q$-factor $Q_\gamma^M(\eta)$ as a linear funtion
         of $\eta$. This line is from $\left(0, \frac{(M-1)\gamma}{
         1+\gamma}\right)$ to $(1,-1)$.} 
\end{figure}

\begin{figure}
\centerline{\epsfxsize=8cm 
\epsfbox{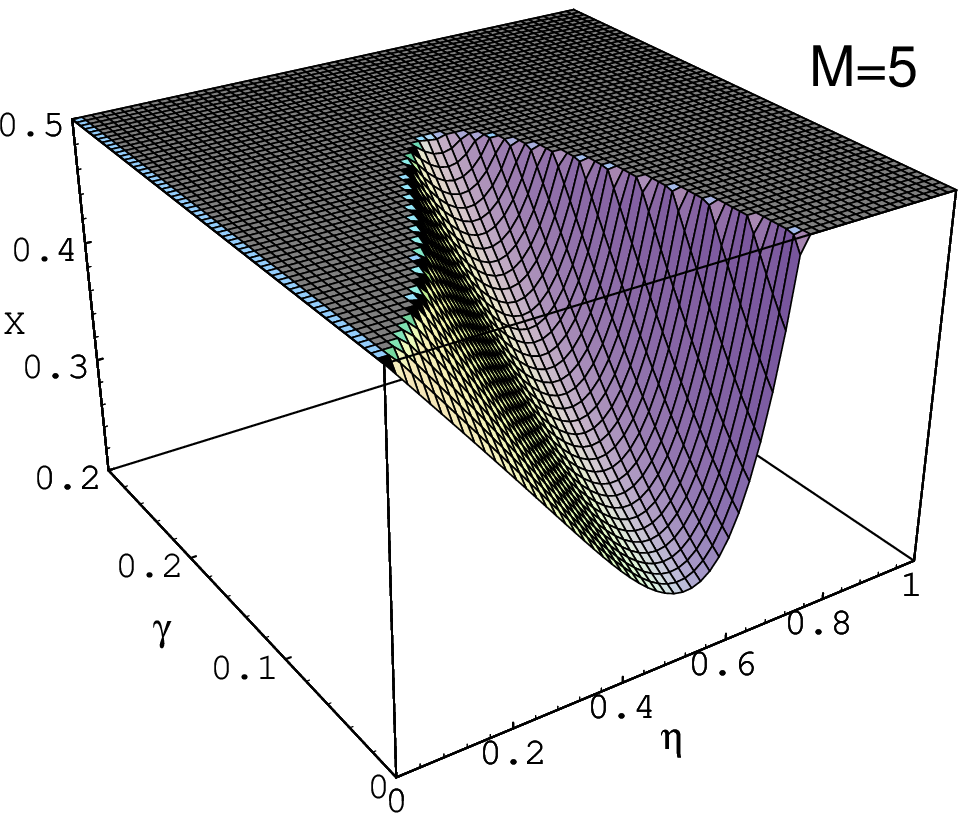}
\epsfxsize=8cm \hspace{1cm} 
\epsfbox{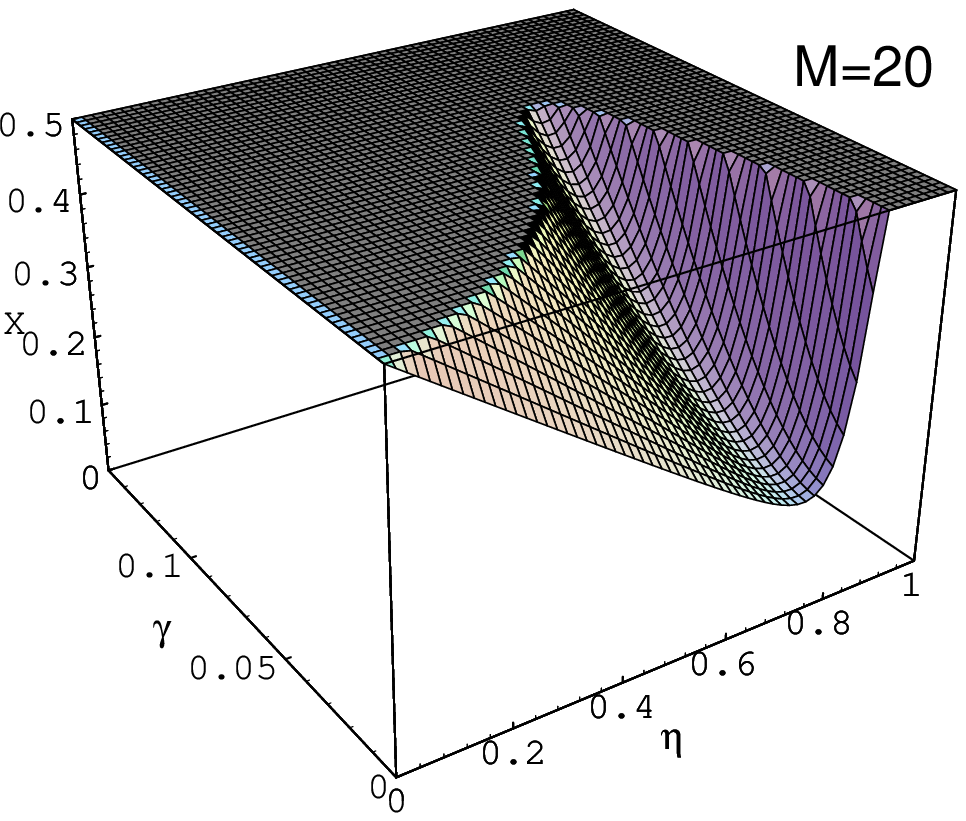}}
\caption{Variance $\langle \Delta x^2\rangle\equiv X$ as a function 
of $\eta$ and $\gamma$ for $M=5,\,20$.} 
\end{figure}

\begin{figure}
\centerline{\epsfxsize=8cm 
\epsfbox{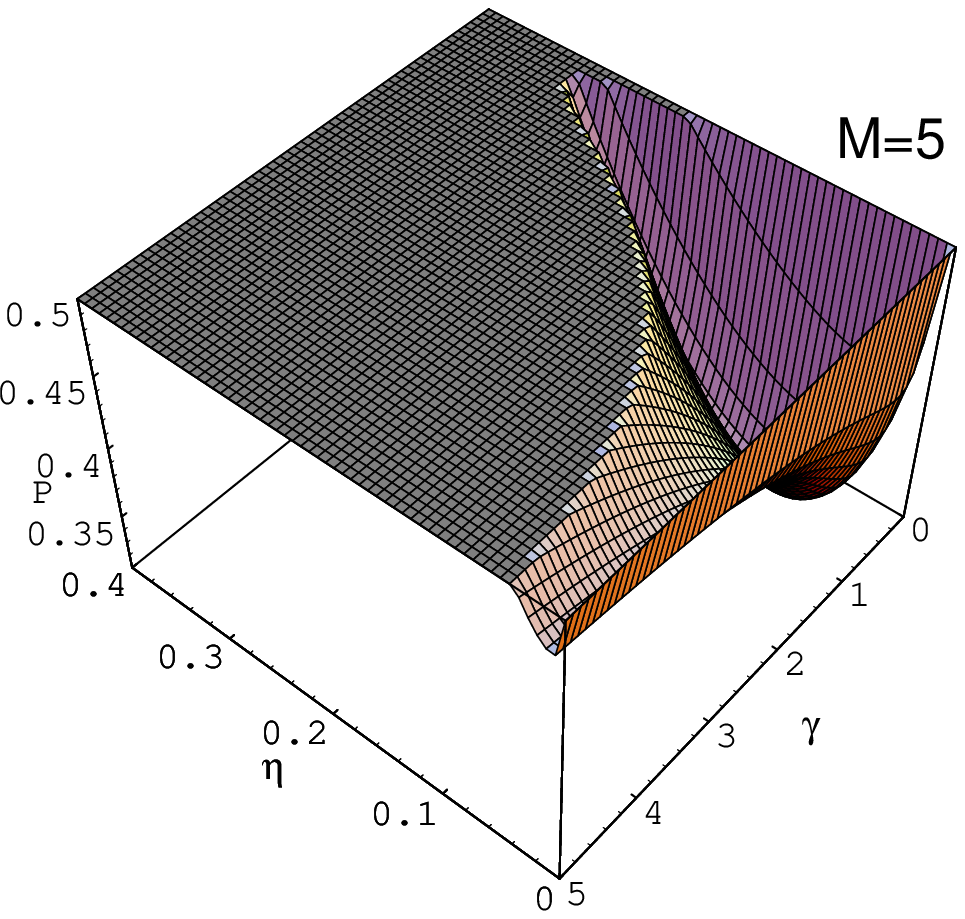}
\epsfxsize=8cm \hspace{1cm} 
\epsfbox{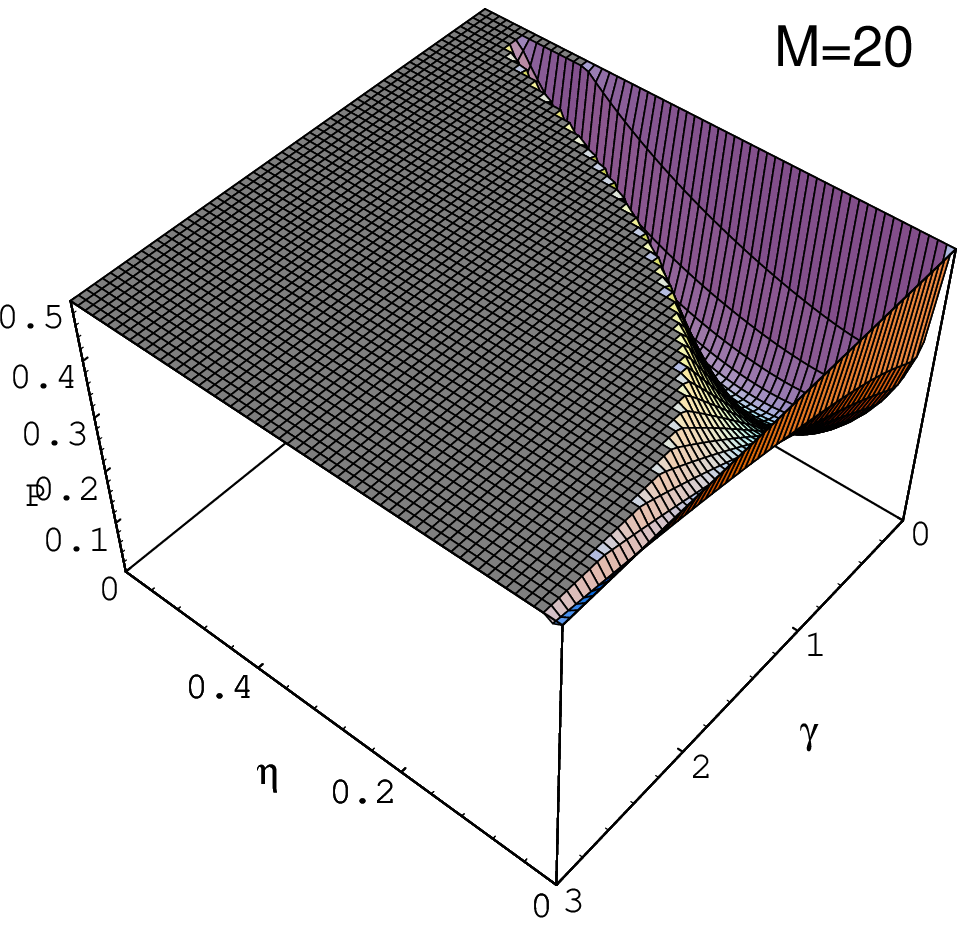}}
\caption{Variance $\langle \Delta p^2\rangle\equiv P$ as a function 
of $\eta$ and $\gamma$ for $M=5,\,20$.}
\end{figure}

\begin{thebibliography}{99}
\bibitem{stol}    Stoler D, Saleh B E A and Teich M C 1985
                  Opt.\,Acta. {\bf 32}, 345
\bibitem{barr}    Vidiella-Barranco A and Roversi J 1994
                  {\it Phys.\,Rev.} {\bf 50A} 5233
\bibitem{nbs}     Fu H C and Sasaki R 1996 {\it Negative Binomial and 
                  Multinomial States: probability distributions and 
                  coherent states} quant-ph/9610022; \\
                  Fu H C and Sasaki R 1996 {\it Negative Binomial 
                  States of quantized radiation fields} quant-ph/9610024 
\bibitem{sg}      Susskind L and Glogower J 1964 {\it Physics} {\bf 1}
                  49
\bibitem{hgs}     Fu H C and Sasaki R 1996 {\it Hypergeometric states
                  and their nonclassical properties} quant-ph/9610021
\bibitem{ins1}    Baseia B, de Lima A F and de Silva A J 1995 {\it
                  Mod.\,Phys.\,Lett.} {\bf 9A} 1673
\bibitem{ins2}    Fu H C and Sasaki R 1996 {\it J.\,Phys.}
                  {\bf 29A} 5637
\bibitem{inp}     Baseia B, de Lima A F and Marques G C 1995 {\it
                  Phys.\,Lett.} {\bf 204A} 1
\bibitem{pro}     Moran P A P 1968 {\it An Introduction to probability
                  theory} (Oxford Science Publications)
\bibitem{pol}     P\'{o}lya G 1930 {\em Annls Inst.\,h.\,Poincar\'{e}.}
                  {\bf 1} 117
\end{thebibliography}
\end{document}